\documentclass{article}
\usepackage{PRIMEarxiv}

\usepackage[utf8]{inputenc} 
\usepackage[T1]{fontenc}    
\usepackage{hyperref}       
\usepackage{url}            
\usepackage{booktabs}       
\usepackage{amsfonts}       
\usepackage{nicefrac}       
\usepackage{microtype}      
\usepackage{lipsum}
\usepackage{fancyhdr}       
\usepackage{graphicx}       
\usepackage[acronyms]{glossaries} 

\usepackage{nccmath}
\usepackage{empheq} 
\usepackage{physics} 
\usepackage{color}
\usepackage{subcaption}
\usepackage{braket}
\usepackage{xcolor}
\usepackage{algorithm}
\usepackage{algorithmicx}
\usepackage{algpseudocode}

\newcommand{\dontshow}[1]{} 

\renewcommand\vec[1]{\overrightarrow{#1}}

\newacronym{cbs}{CBS}{Convergence Born Series}
\newacronym{fft}{FFT}{Fast Fourier Transform}

\makeglossaries
\setkeys{glslink}{hyper=false}  

\title{Iterative Born solver for the Acoustic Helmholtz Equation with Heterogeneous Sound Speed and Density}

\author{
  Antonio Stanziola \\
  Dept. of Medical Physics \& Biomedical Engineering \\
  University College London \\
  London\\
  \texttt{stanziola.antonio@gmail.com} \\
  \And
  Simon R. Arridge \\
  Department of Computer Science \\
  University College London \\
  London\\
  \texttt{s.arridge@ucl.ac.uk} \\
  \And
  Bradley E. Treeby \\
  Dept. of Medical Physics \& Biomedical Engineering \\
  University College London \\
  London\\
  \texttt{b.treeby@ucl.ac.uk} \\
  \And
  Ben T. Cox \\
  Dept. of Medical Physics \& Biomedical Engineering \\
  University College London \\
  London\\
  \texttt{b.cox@ucl.ac.uk} \\
}

\begin{document}
\maketitle

\begin{abstract}
    Efficient numerical solution of the acoustic Helmholtz equation in heterogeneous media remains challenging, particularly for large-scale problems with spatially-varying density—a limitation that restricts applications in biomedical acoustics and seismic imaging. We present a fast iterative solver that extends the Convergent Born Series \cite{osnabrugge2016convergent} method to handle arbitrary variations in sound speed, density, and absorption simultaneously. Our approach reformulates the Helmholtz equation as a first-order system and applies Vettenburg \& Vellekoop's universal split-preconditioner \cite{vettenburg2022universal}, yielding a matrix-free algorithm that leverages Fast Fourier Transforms for computational efficiency. Unlike existing Born series methods, our solver accommodates heterogeneous density without requiring expensive matrix decompositions or pre-processing steps, making it suitable for large-scale 3D problems with minimal memory overhead. The method provides both forward and adjoint solutions, enabling its application for inverse problems. We validate accuracy through comparison against an analytical solution and demonstrate the solver's practical utility through transcranial ultrasound simulations. The solver achieves convergence for strong scattering scenarios, offering a computationally efficient alternative to time-domain methods and matrix-based Helmholtz solvers for applications ranging from medical ultrasound treatment planning to seismic exploration.
\end{abstract}

\keywords{Helmholtz equation \and heterogeneous medium \and Born series \and transcranial ultrasound \and iterative solver \and time harmonic acoustics}

\section{Introduction}

Modeling acoustic wave propagation generated by a single-frequency source through heterogeneous media is essential for many applications, ranging from transcranial ultrasound therapy to seismic imaging, yet remains computationally challenging at high frequencies. In medical ultrasound, for example, accurately simulating waves through the skull—with its high density contrast and complex geometry—requires solving the Helmholtz equation with spatially-varying sound speed, density, and absorption. 

One way to tackle such problems is not to solve the Helmholtz equation directly but to solve the equivalent wave equation in the time domain with a single frequency source, running the simulation for long enough to approximate a steady state, and then extracting the frequency component of interest \cite{aubry2022benchmark}. While this method is simple to setup, it is indirect, computationally intensive, and dispersion errors related to the time discretization can accumulate. An alternative approach recently proposed uses a fixed point scheme for the Helmholtz solution that internally uses a time-domain solver with a temporally-filtered iteration, to reduce the number of timesteps used to reach steady state \cite{appelo2020waveholtz}. 

Classically, techniques based on discretizing the Helmholtz equation and solving the resulting linear system using methods such as iterative, preconditioned, Krylov subspace methods \cite{erlangga2008advances} or multigrid techniques \cite{calandra2013improved} are commonly applied. However, naive applications of these methods often suffer from slow convergence rates and large memory requirements, particularly for problems with high wavenumbers and heterogeneous media. Moreover, finding a preconditioner to improve the convergence rate, for which there is a large literature \cite{gander2019class}, can itself be an expensive task, although if the same problem is to be solved multiple times, it may be worthwhile.
While machine learning approaches have shown promise for solving the Helmholtz equation (e.g. \cite{stanziola2021helmholtz,stanziola2023learned,cui2024neural,veerababu2025solving}), they typically require extensive training data and lack the guaranteed convergence and physical interpretability needed for many scientific applications.

Recently, fast solvers based on the Born series have gained attention due to their ability to efficiently handle large-scale Helmholtz problems. The Born series expresses the solution as the sum of scattering terms, which can be efficiently computed using \glspl{fft} \cite{osnabrugge2016convergent}. However, the classical formulation of the Born series can only be applied
to weakly scattering problems with localized heterogeneity, and quickly diverges for practical problems such as the one encountered in biomedical applications. The \gls{cbs} method \cite{osnabrugge2016convergent} is a preconditioned formulation of the Born series with guaranteed convergence for arbitrarily scattering media, and has demonstrated promising results for solving the heterogeneous Helmholtz equation, with several applications in optics. However, the inability to account for variable density hampers its applicability to real-world acoustic problems where density variations play a significant role in wave propagation, such as in treatment planning in transcranial ultrasonic neurostimulation \cite{marquet2009non}.
To address the limitation of the \gls{cbs} method in handling variable density, several approaches based on matrix approximations have recently been proposed.
One approach in this direction is the use of Homotopy theory and Hierarchical Matrix decomposition \cite{xiang2021homotopy}, which assumes an approximately low-rank structure of the Green's operator to efficiently invert it. 
However, these methods require expensive matrix decomposition or preconditioner construction, similar to LU factorization \cite{ali2024efficient} or randomized matrix approaches \cite{10.1093/gji/ggaa503}. This computational and memory overhead, combined with implementation complexity, limits their scalability to the large 3D problems common in practical applications.

In this paper, we present a novel fast iterative solver for the first-order system of acoustic equations that overcomes the limitations of previous approaches. Our key insight is to reformulate the heterogeneous Helmholtz equation using the universal split-preconditioner \cite{vettenburg2022universal}, which enables convergent iteration even for strong density contrasts. By working with the first-order formulation, we naturally obtain both pressure and velocity fields—essential for computing acoustic intensity in therapeutic applications. The resulting algorithm is matrix-free, requires no preprocessing, and leads to a computationally efficient scheme that can tackle large-scale problems with low memory requirements via the use of efficient FFTs operations. We demonstrate convergence for high sound speed and density contrasts.

\section{Methods}

\subsection{The heterogeneous Helmholtz equation}
The Helmholtz equation describes single-frequency, steady state, wave fields. To derive the version often used in acoustics, we begin with the linearized (time-domain) wave equation in the form
\begin{equation}
    \left(\nabla^2 - \frac{1}{\rho_0}\nabla \rho_0 \cdot \nabla - \frac{1}{c_0^2}\partial_t^2 + \frac{2 \alpha}{c_0} \partial_t  \right)P = S,
    \label{eq:wave_equation}
\end{equation}
where $c_0$, $\rho_0$, and $\alpha$ are the sound speed, mass density, and absorption coefficient, $P$ is the acoustic pressure, and $S$ a source term.
To obtain time harmonic solutions we substitute $P=p e^{-i\omega t}$, and $S = s e^{-i\omega t}$, with $p(x), s(x) \in \mathbb{C}$, where $\omega$ is the circular frequency, giving the Helmholtz equation

\begin{equation}
    \left(\nabla^2 - \frac{1}{\rho_0}\nabla \rho_0 \cdot \nabla + \frac{\omega^2}{c_0^2} - \frac{2 i \omega\alpha}{c_0}  \right)p = s,
\end{equation}

where $p$ is known as the complex acoustic pressure. 
This equation can be written more simply as

\begin{equation}
    \left(\nabla^2 - \frac{1}{\rho_0} \nabla \rho_0 \cdot \nabla + \frac{\omega^2}{c^2}\right)p = s,
    \label{eq:helmholtz1}
\end{equation}
where we have introduced a complex sound speed $c^2(x)$ given by

\begin{equation}
    \label{eq:complex_sound_speed}
    c^2 = \frac{c_0^2}{1 - 2i\alpha c_0 / \omega}.
\end{equation}
(Under the condition that $\alpha \ll \omega / c_0$, this becomes $c \approx c_0 + i\alpha c_0^2/ \omega$, although we will not make use of this approximation in this work.)
This form of the Helmholtz equation with heterogeneous density can be directly discretized in space and, in principle, solved using any technique for linear system inversion. 
Alternatively, instead of starting from the second-order wave equation in time, we can start with the equivalent system of first-order equations \cite{tabei2002k}, which in the frequency domain can be written

\begin{equation}
  \label{eq:Helmholtz-system}
\left\{
\begin{aligned}
  i\omega \mathbf{u} +\rho_0^{-1}\nabla p + \gamma \mathbf{u} &= \mathbf{s}_{\rm u} \\
  i\omega p +\rho_0c^2\nabla \cdot \mathbf{u} +\gamma p &= s_{\rm p} 
\end{aligned}
\right. 
.
\end{equation}

The first-order formulation offers two key advantages over the second-order Helmholtz equation: it allows for an easy rearrangment that naturally separates the material properties (appearing as diagonal terms) from the differential operators (appearing as off-diagonal terms), and it directly provides both pressure and velocity fields useful for intensity calculations. However, it requires $D+1$ complex fields to be stored in memory, where $D \in \{1,2,3\}$ is the dimensionality of the problem.

Bold characters are used throughout to indicate vector fields, eg. the complex acoustic fluid velocity, sometimes called `\textit{particle velocity}', is written as $\mathbf u = \left\{u_x, u_y, u_z\right\}$. The real field $\gamma$ is used to enforce an absorbing layer at the boundary to implement the free-space radiation conditions, and $s_{\rm p}$ and $\mathbf s_{\rm u}$ are complex pressure and velocity source terms. In matrix form this is

\begin{equation}
    \begin{pmatrix}
        i\omega + \boldsymbol{\gamma} & \rho_0^{-1}\nabla \\
        \rho_0c^2\nabla \cdot & i\omega + \gamma 
    \end{pmatrix}
    \begin{pmatrix}
        \mathbf{u} \\
        p \\
    \end{pmatrix}
    =
    \begin{pmatrix}
        \mathbf{s}_{\rm u} \\
        s_{\rm p}
    \end{pmatrix}.
\end{equation}

Since $\rho_0, c^2 \neq 0$, we can make the differential operators homogeneous by multiplying the first row by $\rho_0$ and dividing the second row by $\rho_0c^2$:

\begin{equation}\label{eq:HH1stOrderSH}
    \begin{pmatrix}
        \rho_0(\boldsymbol{\gamma} + i\omega) & \nabla \\
        \nabla \cdot & \frac{\gamma + i\omega}{\rho_0c^2}
    \end{pmatrix}
    \begin{pmatrix}
        \mathbf{u} \\
        p \\
    \end{pmatrix}
    =
    \begin{pmatrix}
       \mathbf{\hat s}_{\rm u}  \\
       \hat s_{\rm p}
    \end{pmatrix},
\end{equation}

where to ease the notation we have defined $\mathbf{\hat s}_{\rm u} = \rho_0\mathbf{s}_{\rm u}$ and $\hat s_{\rm p} = s_{\rm p}/\rho_0c_0^2$.  
This system of equations is equivalent to \eqref{eq:helmholtz1} and, again, can be solved using any linear system inversion technique.

Throughout this work, we assume free-space radiation conditions, implemented via the absorbing boundary layer term $\gamma$. This assumption is well-justified in biomedical acoustics, where acoustic waves typically attenuate before significant reflections occur, and in experimental settings where time-gating can eliminate reflections. While this limits the method to open-domain problems, it enables efficient FFT-based implementation for numerically evaluating the action of the Green's operator, without special boundary handling.

Having established both formulations of the acoustic Helmholtz equation, we now introduce the universal split-preconditioner method that enables efficient iterative solution of the first-order system.

\subsection{Universal split-preconditioner method}

This section will briefly review the universal split-preconditioner method described in \cite{vettenburg2022universal}, of which both our proposed solver and the \gls{cbs} technique are instances. The fundamental idea underlying the Born series is to solve the problem $Ax=y$ by preconditioning it with a suitable preconditioner $\Gamma^{-1}Ax=\Gamma^{-1}y$, for some square invertible operator $\Gamma$, reformulate it into the equivalent equation $x = \Gamma^{-1}y + (I -\Gamma^{-1}A)x$, and then solve using the Neumann series
\begin{align}
    x^* &= \sum_{n=0}^\infty (I - \Gamma^{-1}A)^n\Gamma^{-1}y \nonumber \\
    &= \Gamma^{-1}\left(I + A\Gamma^{-1} + \ldots (A\Gamma^{-1}  )^n + \ldots \right)y ,
    \label{eq:neumann_series}
\end{align}
which converges provided that the spectral radius $(I-\Gamma^{-1}A)$ is smaller than 1. This split-preconditioned method only applies to accretive linear operators. An accretive operator is a linear map that obeys the property

\begin{equation}
    \|(\lambda I + A)x\| \geq \lambda \|x\|,
\end{equation}

for all positive $\lambda$ and for all $x$ in $D(A)$, the domain of $A$. Equivalently it can be defined as

\begin{equation}
    \Re\Braket{x,Ax} \geq 0, \qquad \forall x \in D(A),
\end{equation}

where the notation $\Braket{x,y}$ denotes the inner product. Importantly, skew-Hermitian  systems  are accretive operators \cite{vettenburg2022universal}. This method provides a recipe to find a preconditioner $\Gamma$, using a splitting of the forward operator $A = L + V$, where $V$ is diagonal. When the accretive requirement is satisfied, and assuming that $\|V\| < 1$, it is possible to write the preconditioner

\begin{equation}
    \Gamma^{-1} = \nu^{-1}(I - V)(L + I)^{-1},
\end{equation}
with $0 < \nu < 1$. 
Note that this practically requires to be able to efficiently compute the action of $(L + I)^{-1}$ on vectors. The condition $\|V\| < 1$ can always be satisfied by dividing the entire systems by a large enough constant $c \in \mathbb{C} \setminus \{0\}$, that is 

\begin{equation}
    Ax = y \quad \rightarrow \quad  (A/c)x = y/c
\end{equation}

This preconditioner guarantees that the Neumann series in eq. \eqref{eq:neumann_series} converges to a solution of $Ax = y$. Interestingly, and in line with the classical Born series, it is possible to avoid the application of the forward map $A$, which is the Helmholtz operator in our case. In fact, by rearranging terms, the problem can also be iteratively solved as

\begin{equation}\label{eq:BornAlt}
    x \leftarrow x + \nu B\left[(L+I)^{-1}(Bx + y) - x\right],
\end{equation}
with $B = I - V$. Note that this effectively generalized the Born series technique, and in particular the \gls{cbs} method, as shown in \cite{vettenburg2022universal}.

\subsection{Application to the heterogeneous density case}
To apply this method to our problem, we first need to decouple the differential operators from the diagonal terms

\begin{align}
    A &= 
    \begin{pmatrix}
        \rho_0(i\omega + \boldsymbol{\gamma}) & \nabla \\
        \nabla \cdot & \frac{i\omega + \gamma}{\rho_0c^2}
    \end{pmatrix} \\
    &=
    \begin{pmatrix}
        0 & \nabla \\
        \nabla \cdot & 0
    \end{pmatrix}
    +
    \begin{pmatrix}
        \rho_0(i\omega + \boldsymbol{\gamma}) & 0 \\
        0 & \frac{i\omega + \gamma}{\rho_0c^2}
    \end{pmatrix}.
    \label{eq:time-harmonic-original-system}
\end{align}

Now, note that the diagonal operator is accretive. For the first diagonal element, this is obvious since $\omega, \gamma,\rho_0$ are all positive, so its real part is positive. For the second one, note that it can be written as

\begin{equation}
    \frac{i\omega + \gamma}{\rho_0c^2} =\frac{1}{\rho_0c_0^2}(i\omega + \gamma)(1 - 2i\omega\alpha_0c_0).
\end{equation}

Because $c_0>0$, this is the product of a complex number in the first quadrant with one in the fourth quadrant, and therefore it will have positive real part. To verify also the differential anti-diagonal operator $L$ fulfills the accretive requirement, we need to show that negation gives the adjoint; i.e., for any scalar functions $u, v$ and vector functions ${\bf a}, {\bf b}$ it holds
\begin{equation}
    \left<      \begin{pmatrix} {\bf b} \\ v  \end{pmatrix}, 
\underbrace{\begin{pmatrix}
   0 & \nabla \\
        \nabla \cdot & 0
    \end{pmatrix}}_{L}
    \begin{pmatrix} {\bf a} \\ u  \end{pmatrix}
    \right> = 
    -\left<     \begin{pmatrix} {\bf a} \\ u  \end{pmatrix} ,
\underbrace{\begin{pmatrix}
        0 & \nabla \\
        \nabla \cdot & 0
    \end{pmatrix}}_{-L^{H}}  \begin{pmatrix} {\bf b} \\ v  \end{pmatrix} \right>
\end{equation}

This implies that the operator is skew-Hermitian and, hence, accretive. Since the sum of two accretive operators is still accretive \cite{vettenburg2022universal}, the entire operator A is accretive and we can apply the iterative method. Note that the adjoint operator $A^\mathsf{H}$ satisfies
\begin{align}
    A^\mathsf{H} = &-\begin{pmatrix}
        0 & \nabla \\
        \nabla \cdot & 0
    \end{pmatrix} \\
    &+ \begin{pmatrix}
        \rho_0(-i\omega + \boldsymbol{\gamma}) & 0 \\
        0 & \frac{1}{c_0^2}(-i\omega + \gamma)(1 + 2i\omega\alpha_0c_0)
    \end{pmatrix}.
\end{align}

Following the same arguments as before, as long as $\alpha_0 \geq 0$ the operator is accretive and we can also use the split field preconditioner to find solutions to the adjoint equations. This makes it possible to apply the solver discussed in this paper for inverse problems, such as full waveform inversion imaging \cite{virieux2009overview}, or for embedding physical simulations into trainable machine learning models \cite{karniadakis2021physics}.

\subsubsection{Numerical implementation}
To implement the iterative solver, we must specify the operators $L$ and $V$, choose appropriate scaling parameters, and define a method to compute $(L+I)^{-1}$ efficiently.
Following the approach in \cite{vettenburg2022universal}, we first define the operators $L$ and $V$ as

\begin{align} \label{eq:LV}
    L &= C^{-\frac{1}{2}}\begin{pmatrix}
        a_1 & \nabla \\
        \nabla \cdot & a_2
    \end{pmatrix}C^{-\frac{1}{2}} \\
    V &= C^{-\frac{1}{2}}\begin{pmatrix}
        \rho_0(i\omega + \boldsymbol{\gamma}) - a_1 & 0 \\
        0 & \frac{i\omega + \gamma}{\rho_0c^2} - a_2
    \end{pmatrix}C^{-\frac{1}{2}} \nonumber \\
    &= C^{-\frac{1}{2}}V_0C^{-\frac{1}{2}},
\end{align}

where the diagonal scaling matrix is given by

\begin{align} 
    C &= \begin{pmatrix}
        \lambda_1 & 0 \\
        0 & \lambda_2
    \end{pmatrix} \label{eq:C} \\
    &= 
    \frac{1}{\beta} \begin{pmatrix}
        \max \big[|\rho_0(i\omega + \boldsymbol{\gamma}) - a_a|\big] & 0 \\
        0 & \max \Big[\big|\frac{i\omega + \gamma}{\rho_0c^2} - a_2\big|\Big]
    \end{pmatrix}. \nonumber
\end{align}

The shifts $a_1$ and $a_2$ are choosen to minimize the norm of $V_0$, while the scaling factors $\lambda_1$ and $\lambda_2$ are chosen to ensure that the spectral norm of $V$ is less than 1, e.g. by setting $\beta < 1$, which is a necessary condition for the convergence of the iterative solver.
Similarly we redefine the source term and unknown as
\begin{equation}
    x = C^{\frac{1}{2}}
    \begin{pmatrix}
        \mathbf{u} \\
        p
    \end{pmatrix}, \qquad
    y =  C^{-\frac{1}{2}}
    \begin{pmatrix}
       \mathbf{\hat s}_u  \\
       \hat s_p
    \end{pmatrix}
\end{equation}

The final step in setting up the numerical operators is to calculate the inverse of $(L+I)$, which can be efficiently done in the Fourier domain due to the spatial equivariance of the operator. 
We discretize the velocity field components on a staggered grid, which preserves the skew-Hermitian structure of the differential operators at the discrete level. This choice ensures numerical stability and, as we have empirically verified, provides better convergence to analytical solutions compared to collocated grids. The explicit expressions for $(L+I)^{-1}$ in 2D and 3D, and the details of the staggered grid formulation, are provided in Appendix \ref{sec:numerical_matrices}.

An amplitude scaling factor of $2c_0/dx$ is included to account for the fact that a source when defined on a single grid point is - because this is a spectral model - really a \textit{sinc} function centred on the grid point and zero at all other grid points. The scaling factor can be found by convolving a \textit{sinc} function with the 1-D Green's function for the wave equation.

In short, the iterative computation of the acoustic field using \eqref{eq:BornAlt} with \eqref{eq:LV} and \eqref{eq:C}, is computationally efficient, only involving diagonal matrices and known, simple, operators applied in the Fourier domain using FFTs.

\subsubsection{Practical Algorithm}

The complete iterative algorithm for solving the acoustic Helmholtz equation is:

\begin{algorithm}
\caption{Iterative Solver for Acoustic Helmholtz Equation with Heterogeneous Density}
\begin{algorithmic}
\State \textbf{Input:} Material properties $\rho_0$, $c_0$, $\alpha$; frequency $\omega$; sources $\mathbf{s}_u$, $s_p$
\State \textbf{Preprocessing:}
\State \quad Compute complex sound speed: $c^2 = c_0^2/(1 - 2i\alpha c_0/\omega)$
\State \quad Choose complex shifts, e.g.: $a_1 = \text{median}[\rho_0(i\omega + \gamma)]$, $a_2 = \text{median}[(i\omega + \gamma)/\rho_0c^2]$
\State \quad Compute scaling matrix $C$ using equation \eqref{eq:C}
\State \quad Precompute $(L+I)^{-1}$ in Fourier domain (see Appendix \ref{sec:numerical_matrices})
\State \textbf{Iteration:}
\State \quad Initialize: $x^{(0)} = 0$
\While{$\|x^{(k+1)} - x^{(k)}\| / \|x^{(k)}\| > \text{tolerance}$}
    \State \quad $z = (L+I)^{-1}(Bx^{(k)} + y)$ 
    \State \quad $x^{(k+1)} = x^{(k)} + \nu B(z - x^{(k)})$ 
\EndWhile
\State \textbf{Output:} $[\mathbf{u}, p]^T = C^{-1/2} x^{(k+1)}$
\end{algorithmic}
\end{algorithm}





The computational cost is dominated by the FFT operations in computing $(L+I)^{-1}$, requiring 8 three-dimensional FFTs per iteration (4 forward, 4 inverse) for 3D problems. A numerical implementation of the algorithm is available at \url{https://github.com/ucl-bug/born-density-matlab}.
\section{Numerical Experiments}

\begin{figure*}[t]
    \centering
    \includegraphics[width=\textwidth]{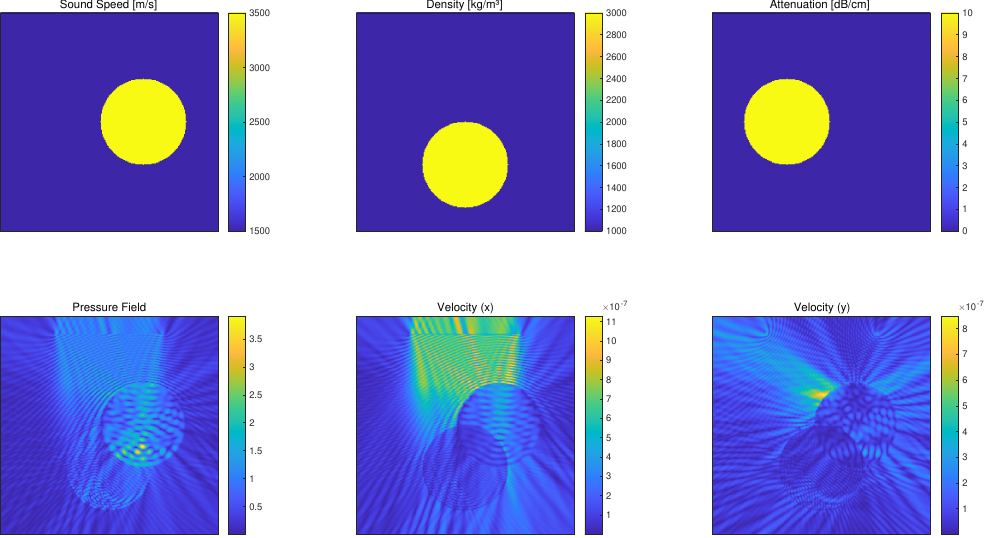}
    \caption{Setup and results for the heterogeneous experiments. The source is a planar line source at the top of the domain.}
    \label{fig:disks_example}
\end{figure*}

To demonstrate the solver's capability in handling complex strongly heterogeneous media, we first present an example featuring distinct localized variations in sound speed, density, and attenuation. The computational domain is a 2D grid of 256$\times$256 points, with a grid resolution of $0.5$mm. The medium properties are defined with a background sound speed of 1500 m/s, which includes a circular inclusion with a higher sound speed of 3500 m/s. Similarly, the background density is set to 1000 kg/m$^3$ and has a circular inclusion with a significantly higher density of 3000 kg/m. For attenuation, the background is set to 0 dB/cm, with a  circular region having an attenuation of 10 dB/cm. The frequency of the problem is set to 500 kHz. The acoustic variables and the resulting acoustic pressure and velocity fields are shown in Fig. \ref{fig:disks_example}, which demonstrate the solver's ability to accurately model wave propagation in highly contrast environments. 

In the following we compare our method to gold-standard computations in two ways : i) for a simple case of scattering from a sphere we compare to an analytical expression; ii) for a more complex heterogeneous domain we compare to a well-established pseudo-spectral time-domain solver.

\subsection{Fluid sphere scattering}

To validate the accuracy of the solver in handling heterogeneous acoustic properties, the example of acoustic scattering from a spherical obstacle was chosen, as Anderson's analytical solution for this case provides an exact solution \cite{anderson1950sound, mcnew2009sound}. The test case consisted of a sphere (radius = 0.4 m) with contrast ratios of 1.5 in both sound speed and density relative to the background medium. A 5 Hz point source was positioned 1.0 meter from the sphere's centre in a cubic domain of 2.1 meters per side. The computational domain was discretized at 24 points per wavelength and surrounded by an absorbing layer spanning 4 wavelengths to minimize boundary reflections. The numerical solution was calculated until convergence to a relative tolerance of $10^{-4}$. The calculations were performed in 3D. The analytic solution was calculated using 150 spherical harmonics; including more harmonics made a negligible difference to the solution.

\begin{figure*}[t]
    \centering
    \includegraphics[width=\textwidth]{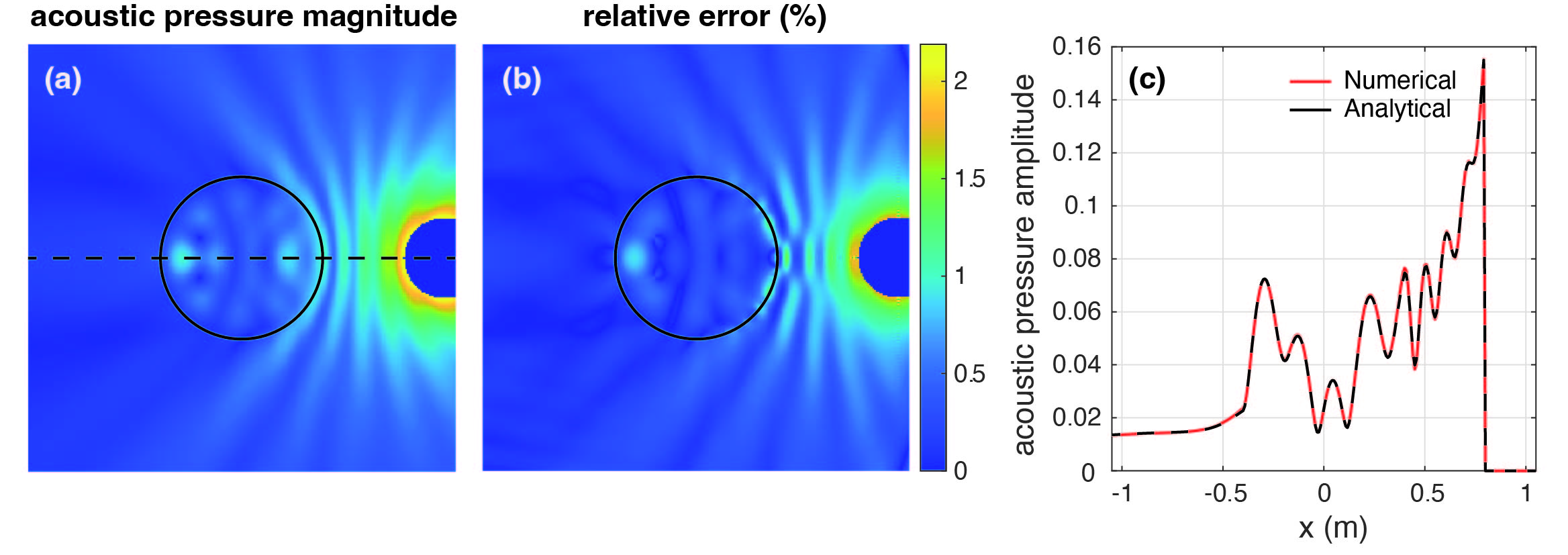}
    \caption{Comparison of the solutions obtained from the proposed numerical solver and an analytic solution obtained using Anderson's spherical harmonic expansion solution\cite{anderson1950sound} for a fluid sphere subject to a point source field. The computations were performed in 3D. (a) Magnitude of acoustic pressure computed using the numerical solver at 24 points per wavelength in the background medium. Central slice shown. (b) Relative error compared to the analytical solution for the same slice. (c) Profiles along the dashed line shown in (a) showing both the numerical and analytical solutions. The region close to the source has been excluded to improve the dynamic range of the plot.}
    \label{fig:afp}
\end{figure*}

Figure \ref{fig:afp} shows a quantitative comparison of the two solutions for a central slice of the domain, excluding the region within one wavelength of the source, as both the truncated expansion of the analytical solution and the grid approximation of the source can't accurately model the proximity of a point source, as its amplitude tends to infinity. 
There is good agreement between the two solutions, confirming the ability of the proposed numerical solver to accurately handle heterogeneous density distributions. The remaining differences are predominantly due to the staircasing effects of modelling the curved surface of a sphere in a rectilinear-grid-based model.

\subsection{Transcranial ultrasound example}
To validate the solver's performance with realistic heterogeneous media, we implemented Benchmark 7 from the international transcranial ultrasound modeling intercomparison study \cite{aubry2022benchmark}. This benchmark represents a clinically relevant scenario of ultrasound propagation through the skull, incorporating strong contrasts in both sound speed and density.

The benchmark consists of a 500 kHz focused bowl transducer (64 mm radius of curvature, 64 mm aperture diameter) targeting the visual cortex through a truncated skull mesh derived from the MNI152 T1 template brain. The transducer was driven with a constant surface velocity of 0.04 m/s, corresponding to a source pressure of 60 kPa in free field. The simulation domain consists of water as the coupling medium and a homogeneous skull model with the following acoustic properties:
\begin{itemize}
    \item Water (background): $c_0$ = 1500 m/s, $\rho_0$ = 1000 kg/m³, $\alpha_0$ = 0 dB/cm
    \item Skull bone: $c_0$ = 2800 m/s, $\rho_0$ = 1850 kg/m³, $\alpha_0$ = 4 dB/cm at 500 kHz
\end{itemize}
These values represent cortical bone properties and create significant impedance mismatches at the water-skull interfaces.

The simulation domain spanned 120 × 70 × 70 mm. Both our solver and the k-Wave reference were run at 12 points per wavelength spatial resolution. An absorbing boundary layer was applied over 50 grid points to implement free-space radiation conditions.  For the k-Wave reference, the simulation was run to steady state using a CFL number of 0.025, ensuring high temporal accuracy. Our iterative solver was run until convergence to a relative tolerance of $2\cdot10^{-5}$.

\begin{figure*}[t]
    \centering
    \includegraphics[width=\textwidth]{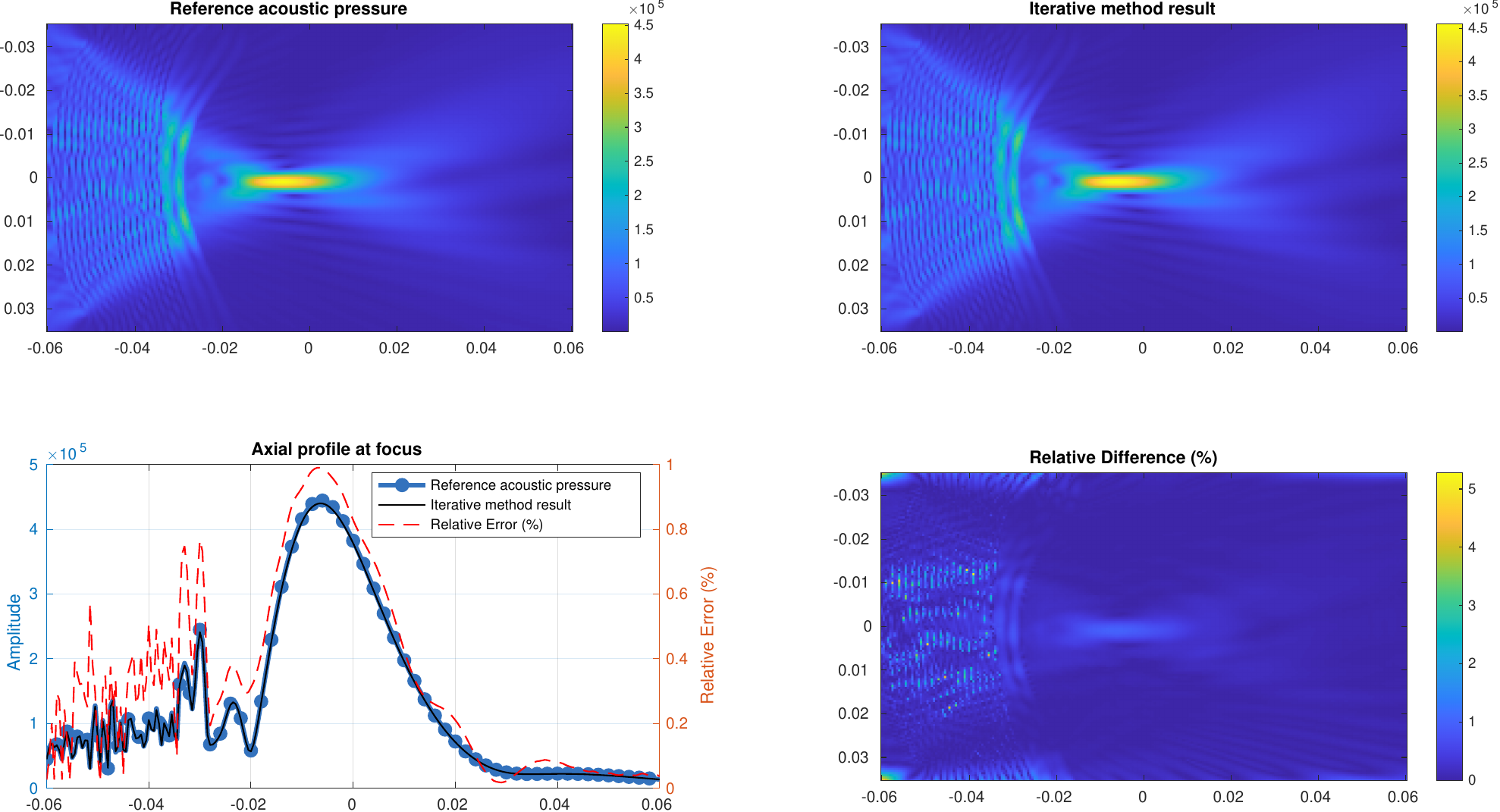}
    \caption{Comparison of the solutions obtained from a reference solve (k-Wave time domain simulation run to steady state) and the proposed numerical solver for a realistic transcranial ultrasound propagation at 500kHz. }
    \label{fig:bm7}
\end{figure*}

The computed pressure fields show excellent agreement with the k-Wave \cite{treeby2012modeling} reference solution. Figure \ref{fig:bm7} displays the pressure amplitude distribution in the central x-y plane for both the reference solver and the proposed solver. The sub-1\% error at the focus—the most clinically relevant metric for therapeutic applications—validates the solver's accuracy for transcranial ultrasound simulations.

\section{Discussion and Conclusions}

We have presented a fast, iterative matrix-free solver for the heterogeneous Helmholtz equation which allows efficient computation of a field in the presence of variable speed, density and attenuation. The method essentially extends the \Gls{cbs} method~\cite{osnabrugge2016convergent} by making use of the universal split preconditioner technique~\cite{vettenburg2022universal} applied to the system of first order equations that decompose the Helmholtz equation.

The proposed solver offers significant practical advantages over existing methods, particularly for large-scale three-dimensional problems involving heterogeneous media. By reformulating the Helmholtz equation into a first-order system and employing a matrix-free approach that leverages FFTs, our method inherently avoids computationally expensive and memory-intensive matrix decompositions or pre-processing steps typically required by direct solvers. This characteristic translates directly into substantial memory savings and faster computation times for large domains, making it highly amenable to modern GPU acceleration. Unlike many existing Born series methods, our approach allows for arbitrary variations in density, which has traditionally been a challenge, extending its application in fields such as biomedical acoustics and seismic imaging where density contrasts are prevalent.

The efficacy and accuracy of our proposed solver have been demonstrated across various challenging scenarios. In a complex transcranial ultrasound simulation, the method achieved sub-1\% error at the focus when compared to established numerical time-domain solutions. This high level of accuracy, coupled with its ability to handle large domains without significant setup time, underscores its potential for realistic applications. In terms of computational complexity, the most expensive operations for our iterative solver are the 3D Discrete Fourier Transforms, implemented using the \gls{fft} algorithm \cite{van1992computational}. For a 3D problem, eight 3-dimensional \glspl{fft} are needed: four forward and four inverse. A possible acceleration of the method could be approached via domain decomposition~\cite{mashe2024}, which allows for coarse-grained parallelisation.

\begin{figure*}[t]
    \centering
    \includegraphics[width=\textwidth]{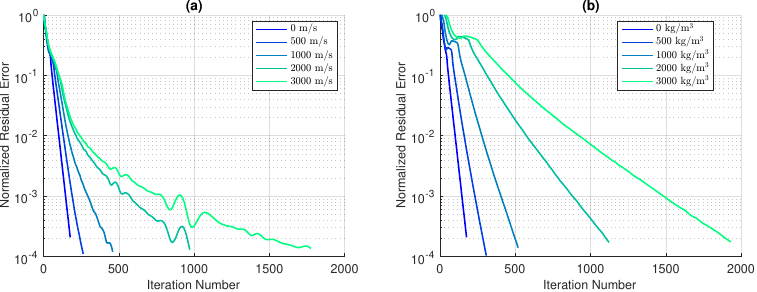}
    \caption{Normalized residual error as a function of iteration count, illustrating the convergence behavior of the iterative solver. (a) Displays convergence for different strengths of sound speed, (b) shows convergence for varying strengths of density. All heterogeneity is introduced as a central disc perturbation of the domain, with background values of $c_0=1500m/s$ and $\rho_0=1000 kg/m^3$. The values in the legend are the additive heterogeneity value, e.g. $500m/s$ indicates an heterogeneity of $1500+500=2000 m/s$ on the sound speed.}
    \label{fig:convergence}
\end{figure*}

Regarding convergence, in line with the \Gls{cbs} method, increasing the sound speed contrast is observed to increase the time to convergence. This occurs as the increased contrast reduces the pseudo-propagation distance of each iteration by reducing the effective support of the $(L+I)^{-1}$ response to a Dirac delta, which has a role analogous to the free-space Green's function in the \Gls{cbs}, impacting the overall speed of the iterative process. An analogous effect is observed for increasing density contrast. An illustration of these qualitative results for a disc inclusion of varying sound speed and density contrast is shown in Fig. \ref{fig:convergence}.

Our method could have applications in ultrasound simulation, especially for applications with high density objects, such as transcranial \cite{tufail2011ultrasonic} and spinal ultrasound simulations \cite{kim2022trans}. This is especially true since it can handle large domains (such as the entire head) without significant setup time and is easy to apply on GPUs. The efficient implementation not just of the direct solver but also its adjoint can be used in the development of algorithms for inverse problems, such as inverse material design and full waveform inversion \cite{plessix2009three}.

Despite its advantages, the proposed matrix-free solver, by operating on a collocation grid, presents certain limitations. While efficient for large domains with distributed heterogeneity, this approach cannot efficiently exploit the sparse structure of problems with localized heterogeneous features. Furthermore, the reliance on \Gls{fft}s for computing the $(L+I)^{-1}$ operator intrinsically links our method to boundary conditions that preserve translation-invariance. This means that directly imposing other common boundary conditions, such as Dirichlet or Neumann conditions on arbitrary surfaces, is not straightforward without fundamentally altering the core computational mechanism. In principle, extending this method to other kinds of simulations, for example on unstructured grids or meshes, would require computing an equivalent free-space Green's function $(L+I)^{-1}$ which is practically localized, making it representable by sparse matrices. However, this would require giving up on its inherent matrix-free approach. For many applications in biomedical acoustics, particularly those involving wave propagation through tissue, the assumption of free-space conditions is often reasonable, as acoustic waves are either strongly attenuated or reflections can be time-gated out during experiments. Alternatively, the method could be extended through neural operators on graphs, which may allow for general geometries and boundary conditions~\cite{li2020neural,lingsch2023beyond}.

A critical consideration for this solver, as in the \Gls{cbs}, is the inherent trade-off in the absorbing layer design. A thin layer requires a high absorption coefficient to be effective against wrap-around artifacts, which can increase the number of iterations required for convergence. Conversely, a thicker absorbing layer, while allowing for weaker absorption and therefore a reduced number of iterations for convergence, increases the computational cost of each iteration due to larger grid dimensions. This trade-off is particularly noticeable in weakly absorbing media. However, in applications where the domain already exhibits significant intrinsic absorption somewhere, the contribution of the absorbing layer to the overall absorption budget is bounded by the largest absorption value in the domain, mitigating this trade-off to some extent.

Another aspect requiring careful consideration is the selection of the optimal values for the parameters $a_1$ and $a_2$ within the universal split preconditioner. While intuition and initial experiments suggest that using the center of the minimum bound circle in the complex domain yields effective results, a comprehensive theoretical framework or a systematic optimization strategy for their general selection across a wider range of heterogeneous media has not been done. Further research into this aspect could lead to improved and more robust convergence characteristics.
\section*{Acknowledgements}
The authors would like acknowledge funding from the UK's Engineering and Physical Sciences Research Council (EPSRC), grant numbers EP/W029324/1 and EP/T014369/1.

\newpage
\appendix
\section{Derivation of the time-harmonic wave equation}
Starting from \eqref{eq:Helmholtz-system}, ignoring the absorbing boundary $\gamma$:
\begin{equation}
\left\{
\begin{aligned}
  i\omega \mathbf{u}  +\rho_0^{-1}\nabla p &= \mathbf{s}_{u } \\
  i\omega p +\rho_0 c^2\nabla \cdot \mathbf{u} &= s_p 
\end{aligned}
\right. 
.
\end{equation}
The first equation gives, componentwise,
\begin{equation}
    u_\eta = -\frac{1}{i\omega}\rho_0^{-1}\partial_\eta p + \frac{s_{u_{\eta}}}{i\omega}.
\end{equation}
Plugging it in the second one 
\begin{equation}
    i\omega p + \rho_0 c^2 \sum_{\eta}\partial_\eta \left[ -\frac{1}{i\omega}\rho_0^{-1}\partial_\eta p + \frac{s_{u_{\eta}}}{i\omega} \right] = s_p;
\end{equation}
multiplying by $-i\omega/c^2$
\begin{equation}
    \frac{\omega^2}{c^2}p + \sum_{\eta}\left[\rho_0 \partial_\eta  \rho_0^{-1}\partial_\eta p - \rho_0\partial_\eta s_{u_{\eta}} \right] = -\frac{i\omega}{c^2}s_p.
\end{equation}
Applying the chain rule and moving all sources to the same side gives
\begin{equation}
    \frac{\omega^2}{c^2}p + \sum_{\eta}\left[ \rho_0 (\partial_\eta \rho_0^{-1})(\partial_\eta p) + \partial^2_\eta p\right] = -\frac{i\omega}{c^2}s_p + \rho_0\sum_{\eta}\partial_\eta s_{u_{\eta}}.
\end{equation}
Using the fact that $\rho_0 \partial_\eta \rho_0^{-1} = - \rho_0^{-1}\partial_\eta \rho_0$, rearranging and writing it in vector notation gives
\begin{equation}
    \left(\nabla^2 - \frac{1}{\rho_0} \nabla \rho_0 \cdot \nabla +  \frac{\omega^2}{c^2}\right)p = s,
        \label{eq:helmholtz2}
\end{equation}
which corresponds to the time harmonic equation when one uses \eqref{eq:complex_sound_speed} as $c^2$. Note that the equivalent pressure source is given by
\begin{equation}
    s = -\frac{i\omega}{c^2}s_p + \rho_0\nabla \cdot \mathbf{s}_u
\end{equation}
therefore $s_p$ is really acting as a sound-speed scaled version of a mass source.
\section{Numerical matrices}
\label{sec:numerical_matrices}
Omitting the forward and inverse Fourier transforms, in 2D and 3D the operator $(L+I)^{-1}$ is

\begin{equation}
    (L+I)^{-1} = \frac{1}{\mu}
    \begin{pmatrix} 
    \dfrac{\lambda_0}{a + \lambda_0}\left(\mu I -K\right) & -i\sqrt{\lambda_0 \lambda_1} {\bf k} \\
    -i\sqrt{\lambda_0 \lambda_1} {\bf k}^{\rm T} & \lambda_1 (a + \lambda_0) 
    \end{pmatrix}
\end{equation}

where $\bf k$ is the k-vector and 

\begin{equation}
    \mu = k^2 + (a + \lambda_0)(b+\lambda_1), \qquad K = {\bf k}\otimes{\bf k}
\end{equation}
For example, in 3D this is equal to

\begin{align}
    &(L+I)^{-1}_{\text{3D}} = \frac{1}{\mu} \times \nonumber \\
    & \begin{pmatrix}
    \dfrac{\lambda_0(\mu - k_x^2)}{a + \lambda_0} &
    -\dfrac{\lambda_0k_x k_y }{a + \lambda_0} &
    -\dfrac{\lambda_0k_x k_z }{a + \lambda_0} &
    -ik_x \sqrt{\lambda_0 \lambda_1} \\
    -\dfrac{\lambda_0k_x k_y }{a + \lambda_0} &
    \dfrac{\lambda_0(\mu - k_y^2)}{a + \lambda_0} &
    -\dfrac{\lambda_0k_y k_z }{a + \lambda_0} &
    -ik_y \sqrt{\lambda_0 \lambda_1} \\
    -\dfrac{\lambda_0k_x k_z }{a + \lambda_0} &
    -\dfrac{\lambda_0k_y k_z }{a + \lambda_0} &
    \dfrac{\lambda_0(\mu - k_z^2)}{a + \lambda_0} &
    -ik_z \sqrt{\lambda_0 \lambda_1} \\
    -ik_x \sqrt{\lambda_0 \lambda_1} &
    -ik_y \sqrt{\lambda_0 \lambda_1} &
    -ik_z \sqrt{\lambda_0 \lambda_1} &
    \lambda_1 (a + \lambda_0)
    \end{pmatrix}. 
\end{align}

In case one needs to solve the adjoint problem, note that we can simply use the element-wise conjugate of the matrices above:

\begin{equation}
    (L^\mathsf{H}+I)^{-1} = \left[ (L+I)^\mathsf{H} \right]^{-1} = \left[ (L+I)^{-1} \right]^\mathsf{H} = \left[ (L+I)^{-1} \right]^*,
\end{equation}
where the last passage uses the fact that $(L+I)^{-1}$ is symmetric.


\subsection{Staggered formulation}
Let us define the shift operator that acts component-wise as

\begin{equation}
    S^+u = \begin{pmatrix}
        S^+_x u \\
        \\
        S^+_y u \\
        \\
        S^+_z u \\
    \end{pmatrix} = \begin{pmatrix}
        u\left(x+\dfrac{\Delta}{2},y,z\right) \\
        \\
        u\left(x,y+\dfrac{\Delta}{2},z\right) \\
        \\
        u\left(x,y,z+\dfrac{\Delta}{2}\right) \\
    \end{pmatrix}, 
    \qquad
    S^+\begin{pmatrix}
        u_x \\
        u_y \\
        u_z
    \end{pmatrix} = \begin{pmatrix}
        S^+_x & 0 & 0 \\
        \\
        0 & S^+_y & 0 \\
        \\
        0 & 0 & S^+_z \\
    \end{pmatrix}\begin{pmatrix}
        u_x \\
        u_y \\
        u_z
    \end{pmatrix} = \begin{pmatrix}
        u_x\left(x+\dfrac{\Delta}{2},y,z\right) \\
        \\
        u_y\left(x,y+\dfrac{\Delta}{2},z\right) \\
        \\
        u_z\left(x,y,z+\dfrac{\Delta}{2}\right) \\
    \end{pmatrix},
\end{equation}
where $\Delta$ is the grid spacing (here assumed uniform, but this can be trivially extended to non-uniform grids).

Its inverse operator $S^-$ works in exactly the same way, with the difference that it subtracts $\Delta/2$, so clearly $S^-S^+ = I$. Note also that they satisfy the property $S^+\left(u^{-1}\right) = \left(S^+u\right)^{-1}$, and the property $S^+(uv) = \left(S^+u\right)\left(S^+ v\right)$.

If we want to evaluate the velocity field components on forward staggered grids, the system in eq. \eqref{eq:time-harmonic-original-system} is now

\begin{equation}
    \begin{pmatrix}
        i\omega + S^+\boldsymbol{\gamma} & S^+(\rho_0^{-1})S^+\nabla  \\
        \rho_0c^2\nabla \cdot S^- & i\omega + \gamma 
    \end{pmatrix}
    \begin{pmatrix}
        \mathbf{u} \\
        p \\
    \end{pmatrix}
    =
    \begin{pmatrix}
        \mathbf{s}_u \\
        s_p
    \end{pmatrix}.
    \label{eq:time-harmonic-original-system-stag}
\end{equation}

Note that we assume that $\mathbf{u}$ and $\mathbf{s}_u$ are already defined on the staggered grid. Doing the same manipulations as before to isolate the differential operators we get

\begin{equation}
    \begin{pmatrix}
        \left(S^+\rho_0\right)\left(S^+\boldsymbol{\gamma} + i\omega\right) & S^+ \nabla  \\
        \nabla \cdot S^- & \frac{\gamma + i\omega}{\rho_0c^2}
    \end{pmatrix}
    \begin{pmatrix}
        \mathbf{u} \\
        p \\
    \end{pmatrix}
    =
    \begin{pmatrix}
       \mathbf{\hat s}_u  \\
       \hat s_p
    \end{pmatrix},
\end{equation}

where now the velocity source was divided component-wise by the staggered density. Splitting the operator now returns

\begin{equation}
    A = \begin{pmatrix}
        0 & S^+ \nabla  \\
        \nabla \cdot S^- & 0
    \end{pmatrix}
    + 
    \begin{pmatrix}
        \left(S^+\rho_0\right)\left(S^+\boldsymbol{\gamma} + i\omega\right) & 0 \\
        0 & \frac{\gamma + i\omega}{\rho_0c^2}
    \end{pmatrix}.
\end{equation}

When discretizing the equation, we now have to define how the shift operators are implemented. With respect to the term $\gamma$ this is trivial, as it is generated algorithmically and therefore its shifted version $\vec{S}\boldsymbol{\gamma} = \vec{\boldsymbol{\gamma}}$ can be easily evaluated at the staggered grid points

However, it is likely that the user will only have $\rho_0$ defined on the unstaggered computational grid. We therefore compute its shifted version using linear interpolation

\begin{equation}
    S^+\rho_0 \simeq \vec{\rho_0} = \frac{1}{2}\begin{pmatrix}
        {\rho_0}(x,y,z) + {\rho_0}(x+\Delta,y,z) \\
        {\rho_0}(x,y,z) + {\rho_0}(x,y+\Delta,z) \\
        {\rho_0}(x,y,z) + {\rho_0}(x,y,z+\Delta)
    \end{pmatrix}.
\end{equation}

For the non-diagonal operator, one can use the shift theorem of the Fourier transform to show that

\begin{equation}
    \partial_\eta S^+_\eta = S^+_\eta \partial_\eta  =\mathcal{F}^\mathsf{H}\left(i k_\eta e^{i k_\eta\Delta/2}\right)\mathcal{F}.
\end{equation}

To simplify the notation, let's define the following discrete shift filters in the Fourier domain:

\begin{align}
    S_\eta &= e^{i k_\eta\Delta/2}, \qquad S^\eta = \left(S_\eta \right)^* = e^{-i k_\eta \Delta/2}, \\
    \qquad S_\eta^\mu &= S_\eta S^\mu = S^\mu S_\eta =e^{i \left(k_\eta - k_\mu\right)\Delta/2}. \nonumber
\end{align}

Clearly, $S^\eta S_\eta = 1$. The operator now becomes

\begin{equation}
    \begin{pmatrix}
        0 & S^+ \nabla  \\
        \nabla \cdot S^- & 0
    \end{pmatrix}
    = \mathcal{F}^\mathsf{H}
        \begin{pmatrix}
        0 & 0 & 0 & i k_x S_x \\
        0 & 0 & 0 & i k_y S_y \\
        0 & 0 & 0 & i k_z S_z \\
        i k_x S^x & i k_y S^y & i k_z S^z & 0
    \end{pmatrix}
    \mathcal{F}.
\end{equation}
The adjoint of this operator is
\begin{align}
   & \begin{pmatrix}
        0 & S^+ \nabla  \\
        \nabla \cdot S^- & 0
    \end{pmatrix}^\mathsf{H} \nonumber \\
    &= \mathcal{F}^\mathsf{H}
        \begin{pmatrix}
        0 & 0 & 0 & -i k_x S_x \\
        0 & 0 & 0 & -i k_y S_y \\
        0 & 0 & 0 & -i k_z S_z \\
        -i k_x S^x & -i k_y S^y & -i k_z S^z & 0
    \end{pmatrix}
    \mathcal{F}
\end{align}
so, once again, this is a skew-Hermitian operator, and therefore we can apply the universal split preconditioning method to the staggered formulation. The required inverse is then

\begin{equation}
    (L+I)^{-1} = \frac{1}{\mu}
    \begin{pmatrix} 
    \dfrac{\lambda_0}{a + \lambda_0}S^+\left(\mu I -K\right)S^- & -i\sqrt{\lambda_0 \lambda_1} {\bf k}S^- \\
    -i\sqrt{\lambda_0 \lambda_1} {\bf k}^{\rm T}S^+ & \lambda_1 (a + \lambda_0) 
    \end{pmatrix}.
\end{equation}
In 3D, for example, this is equal to
\begin{equation}
    (L+I)^{-1}_{\text{3D}} = \frac{1}{\mu}
    \begin{pmatrix}
    \dfrac{\lambda_0(\mu  - k_x^2)}{a + \lambda_0} &
    -\dfrac{\lambda_0k_x k_y S_x^y }{a + \lambda_0} &
    -\dfrac{\lambda_0k_x k_z S_x^z}{a + \lambda_0} &
    -ik_x \sqrt{\lambda_0 \lambda_1} S_x\\
    -\dfrac{\lambda_0k_x k_y S_y^x}{a + \lambda_0} &
    \dfrac{\lambda_0(\mu - k_y^2)}{a + \lambda_0} &
    -\dfrac{\lambda_0k_y k_z S_y^z}{a + \lambda_0} &
    -ik_y \sqrt{\lambda_0 \lambda_1} S_y\\
    -\dfrac{\lambda_0k_x k_z S_z^x}{a + \lambda_0} &
    -\dfrac{\lambda_0k_y k_z S_z^y}{a + \lambda_0} &
    \dfrac{\lambda_0(\mu - k_z^2)}{a + \lambda_0} &
    -ik_z \sqrt{\lambda_0 \lambda_1} S_z\\
    -ik_x \sqrt{\lambda_0 \lambda_1} S^x&
    -ik_y \sqrt{\lambda_0 \lambda_1} S^y&
    -ik_z \sqrt{\lambda_0 \lambda_1} S^z&
    \lambda_1 (a + \lambda_0)
    \end{pmatrix}.
\end{equation}

\section{Analytic solution for the fluid sphere}

This derivation reproduces the results from \cite{anderson1950sound}, but closely follows the derivation in \cite{mcnew2009sound}. Starting form the outer medium, the field is composed by two parts: the incident wave and the scattered wave
$$
p_0 = p_{0i} + p_{0r},
$$
which can be decomposed in spherical harmonics as
$$
p_{0i} = \sum_{m=0}^{\infty} (2m+1)\mathcal{L}_m P_m(\mu) \, j_m(k_0r)
$$
$$
p_{0r} = \sum_{m=0}^{\infty} A_m P_m(\mu) \, h_m(k_0r),
$$
where $P_m$ indicates the Legendre polynomial of degree $m$, $j_m$ is the spherical Bessel function of order $m$, $h_m$ is the spherical Hankel function of order $m$, and $A_m$, $B_m$ are coefficients to be determined. Lastly, the coefficient $\mathcal{L}_m$ depends on the kind of incident wave, and it is $\mathcal{L}_m = h_m(k_0R)$ for a monopole source of unit amplitude, where $R$ is the sphere radius. While it might appear superfluous to write down the incident wave in this way, as of course one can compute it in a single pass for plane and monopole sources, this will prove to be useful to match orders later in the derivation.

The inner region can be modeled as 
$$
p_{1} = \sum_{m=0}^{\infty} B_m P_m(\mu) \, j_m(k_1r)
$$
It is now possible to equate both the radial velocity and the pressure at the interface. For the velocities, one has
$$
u_{0i} = \frac{1}{Z_0} \frac{\partial}{\partial(k_0r)} p_{0i} = \frac{1}{Z_0} \sum_{m=0}^{\infty} (2m+1)\mathcal{L}_m P_m(\mu) \, j'_m(k_0r)
$$
$$
u_{0r} = \frac{1}{Z_0}\sum_{m=0}^{\infty} A_m P_m(\mu) \, h'_m(k_0r)
$$
$$
u_{1} = \frac{1}{Z_1} \sum_{m=0}^{\infty} B_m P_m(\mu) \, j'_m(k_1r).
$$
Here $Z_i$ is the acoustic impedance of the $i$-th region, with $i=0$ outside the sphere and $i=1$ inside. The next step is to use the orthogonality properties of the basis functions used, equating them at each order at the interface $r=R$, resulting in two equations
$$
\frac{(2m+1) \mathcal{L}_m}{Z_0}j'_m(k_0R) + \frac{h'_m(k_0R)}{Z_0}A_m = \frac{j'_m(k_1R)}{Z_1}B_m
$$
$$
(2m+1) \mathcal{L}_mj_m(k_0R) + h_m(k_0R)A_m = j_m(k_1R)B_m
$$
In matrix form, this is
\begin{align*}
&\begin{pmatrix}
h'_m(k_0R)/Z_0 & - j'_m(k_1R)/Z_1 \\
\\
-h_m(k_0R) & j_m(k_1R) \\
\end{pmatrix}
\begin{pmatrix}
A_m \\
B_m
\end{pmatrix} \\
&= (2m+1) \mathcal{L}_m
\begin{pmatrix}
-j'_m(k_0R)/Z_0 \\
\\
j_m(k_0R)
\end{pmatrix}
\end{align*}
The inverse of the matrix is
\begin{align*}
&\dfrac{1}{
-Z_0j'_m(k_1R)h_m(k_0R) + Z_1h'_m(k_0R) j_m(k_1R) } \times \\
&\begin{pmatrix}
Z_0Z_1j_m(k_1R) & Z_0 j'_m(k_1R) \\
\\
Z_0Z_1 h_m(k_0R) & Z_1h'_m(k_0R)
\end{pmatrix}
\end{align*}
Therefore, the two coefficients are given by
\begin{align*}
&A_m = \mathcal{L}_m(2m+1) \times \\
&\dfrac{Z_0j_m(k_0R)j'_m(k_1R)-Z_1j_m(k_1R)j'_m(k_0R)}{
-Z_0j'_m(k_1R)h_m(k_0R) + Z_1h'_m(k_0R) j_m(k_1R) }
\end{align*}
\begin{align*}
&B_m = \mathcal{L}_m(2m+1) \times \\
&\dfrac{-Z_1h_m(k_0R)j'_m(k_0R) + Z_1 h'_m(k_0R)j_m(k_0R)}{
-Z_0j'_m(k_1R)h_m(k_0R) +Z_1h'_m(k_0R) j_m(k_1R)}.
\end{align*}
Note that the coefficient $A_m$ matches the Anderson result as rewritten in \cite{mcnew2009sound}. 

The latter two formulas allow to calculate the value of the filed inside and outside the sphere up to an arbitrary order $m$. In the experiments, $m$ has been set to 150.


\end{document}